\documentclass[prl,preprint,double-spaced,superscriptaddress,preprintnumbers,onecolumn,preprint,aps]{revtex4}
\usepackage{graphicx}
\usepackage{dcolumn}
\usepackage{bm}

\begin{document}

\title{Charge-Stripe Order in a Parent Compound of Iron-based Superconductors}

\author
{Wei Li, $^{1,2}$ Wei-Guo Yin, $^{3\ast}$ Lili Wang, $^{1,2}$ Ke He, $^{1,2}$ Xucun Ma, $^{1,2}$ \\ Qi-Kun Xue, $^{1,2\ast}$ Xi Chen $^{1,2\ast}$\\
\normalsize{$^{1}$State Key Laboratory of Low-Dimensional Quantum Physics,}\\
\normalsize{Department of Physics, Tsinghua University, Beijing 100084, China}\\
\normalsize{$^{2}$Collaborative Innovation Center of Quantum Matter, Beijing 100084, China}\\
\normalsize{$^{3}$Condensed Matter Physics and Materials Science Department,}\\
\normalsize{Brookhaven National Laboratory, Upton, New York 11973, USA}
\\
\normalsize{$^\ast$To whom correspondence should be addressed. E-mail:  wyin@bnl.gov (W. G. Y);}\\
\normalsize{ qkxue@mail.tsinghua.edu.cn (Q. K. X.); xc@mail.tsinghua.edu.cn (X. C.).}
}

\begin{abstract}
Charge ordering is one of the most intriguing and extensively studied phenomena in correlated electronic materials because of its strong impact on electron transport properties including superconductivity.
Despite its ubiquitousness in correlated systems, the occurrence of charge ordering in iron-based superconductors is still unresolved. Here we use scanning tunneling microscopy to reveal a long-range charge-stripe order and a highly anisotropic dispersion of electronic states in the ground state of stoichiometric FeTe, the parent compound of the Fe(Te, Se, S) superconductor family.
The formation of charge order in a strongly correlated electron system with integer nominal valence (here Fe$^{2+}$) is unexpected and suggests that the iron-based superconductors may exhibit more complex charge dynamics than originally expected. We show that the present observations can be attributed to the surpassing of the role of local Coulomb interaction by the poorly screened longer-range Coulomb interactions, facilitated by large Hund's rule coupling.

\end{abstract}

\maketitle

Charge order (CO) has been observed in a wide range of  strongly correlated electron systems (SCESs), such as manganites \cite{dagotto05,coey04}, magnetite \cite{senn12}, cobaltates \cite{cwik09}, nickelates \cite{billinge13}, and cuprates \cite{uchida95,lawler10,comin14,neto14}.
Such ubiquity comes unexpected since the two generic leading energy scales in electron systems---the kinetic energy and local Coulomb interaction $U$---generally favor uniform charge distribution instead.
However, it has been shown that CO may result from a compromised interplay of charge with the spin~\cite{gunnarsson89} and/or orbital~\cite{Volja,Brink} degrees of freedom. Moreover, the bad-metal behavior of SCESs \cite{Emery,note:bad-metal} implies that CO can also be driven by the poorly screened longer-range Coulomb interactions. 
Usually, the SCESs that exhibit CO have fractional nominal valence due to charge doping or mixed valence in nature \cite{note:valence}, which allows charge fluctuation free of the energy penalty from $U$. An exception comes when $U$ is surpassed by Hund's rule coupling \cite{Mazin}. The ubiquity of CO is thus a manifestation of electronic correlation effects \cite{dagotto05}.
Experimentally, CO is readily characterized by diffraction, x-ray scattering, and scanning tunneling microscopy (STM) techniques \cite{uchida95,lawler10,comin14,neto14}.
These features of CO have made its study an effective route to understanding SCESs in general and helping resolve some current grand problems in condensed matter physics, such as high-temperature superconductivity in cuprates and colossal magnetoresistance in manganites in particular.

Naturally, it is desirable to know how CO emerges in iron-based superconductors (FeSCs) \cite{hosono08},  which appear to have all the aforementioned ingredients for CO: they contain the charge, spin, orbital degrees of freedom \cite{yin09} and show bad-metal behavior \cite{si08}. Oddly enough, to date, convincing evidence of CO in FeSCs is still lacking.  Compared with the other SCESs, FeSCs are indeed somehow different. For example, the origin of strong electron correlation in FeSCs is attributed to the Hund's rule coupling instead of $U$ \cite{medici14,yin11_NM,yin10,yin12} likely due to the suppression of $U$ via strongly
coupled charge multipole polarizations of Fe and anion  \cite{ma14}. This peculiar charge dynamics could mediate electron pairing for superconductivity \cite{sawatzky09}. On the other hand, the suppression of $U$ also favor the formation of CO.
Hence, finding CO in FeSCs has become an important step to our understanding of FeSCs and CO mechanisms in SCESs.

Here we report the direct observation of long-range static charge order in FeSCs with systematic STM measurement on FeTe, the parent compound of the Fe(Te,Se,S) family of FeSCs \cite{wu08,gu10,bao09,dai09,xiang09}.  FeTe was chosen because of its simple structure (Fig.~1A) and arguably strongest electronic correlation in the FeSC system \cite{yin11_NM}. The ground state of bulk FeTe is known to possess the metallic \emph{bicollinear} antiferromagnetic (AFM) spin order \cite{bao09,dai09,xiang09}, named after the pattern of alternating two columns of Fe sites with spin-up and two columns of Fe sites with spin-down (see Fig.~1B). Alternatively, the two columns of Fe sites of the same spin can be viewed as a zigzag chain formed by the nearest Fe-Fe bonds; this AFM order is called \emph{$E$-type} in the context of manganites \cite{yin10}. The latter viewpoint bridges these two important classes of SCESs: FeSCs and manganites. In fact, it has been shown that various material-dependent magnetic orders in FeSCs \cite{yin10,yin12} and manganites \cite{hotta03} are unified in a spin-fermion model.
In this work, we find that CO in FeTe peculiarly follows the same bicollinear $E$-type order, a pattern that has not been observed or predicted before in the charge channel. Even more strikingly, this CO is realized in a SCES with integer nominal valence. We suggest that the $E$-type CO can be well accounted for by including long-range Coulomb interaction in the spin-fermion model.

The experiments were carried out in a Unisoku UHV $^3$He STM system equipped with a molecular beam epitaxy (MBE) chamber for {\it in situ} film growth. Single crystal Fe$_{1+x}$Te samples usually contain a sizeable amount of excessive Fe atoms, which are known to bring substantial extrinsic effects such as transforming the metallic FeTe to a semiconductor \cite{bao09,note:extrinsic}.
To avoid these extrinsic effects, we grew stoichiometric FeTe single-crystalline films with MBE
in ultra-high vacuum (UHV) and performed the STM experiment in the same UHV system \cite{SI}.
Topography of the Te-terminated FeTe film (Fig. 1C) shows the atomically flat surface with broad terraces.
The step height is 0.63 nm. The image with atomic resolution  (Fig. 1C, inset) exhibits a quasi-square lattice of Te atoms
with lattice constant of $\sim$ 3.8 \AA.
Previous studies \cite{bao09,dai09} on bulk FeTe show a tetragonal to monoclinic structural phase transition at $T_\mathrm{s} \sim 65$~K
with a simultaneous development of the $E$-type AFM order (i.e., $T_\mathrm{N}=T_\mathrm{s}$).

\begin{figure}[t]
        \includegraphics[width=6in]{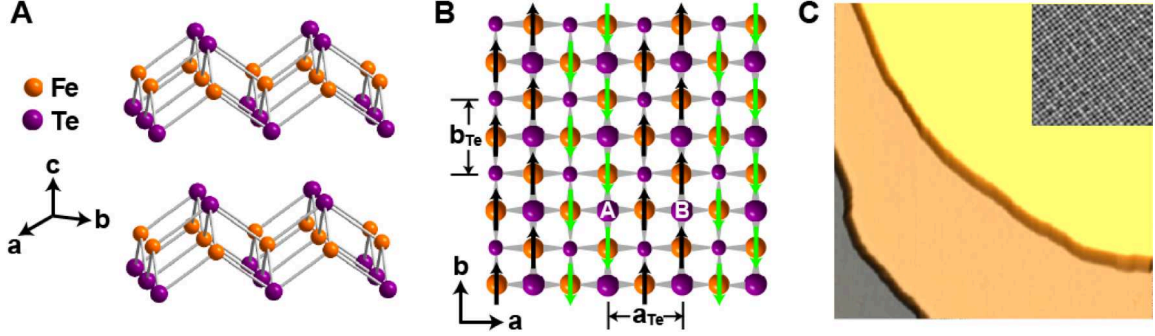}
        \caption{\label{fig1} Structure and MBE film of FeTe. (A) Crystal structure of FeTe. (B) Bicollinear spin structure of FeTe.
        The arrows indicate the spin orientations of iron atoms.
        The smaller balls indicate the second-layer Te atoms.
        (C) Topography of stoichiometric (no excess Fe) FeTe(001) film (3.3 V, 0.03 nA, 200 nm $\times$ 200 nm). The inset shows the square lattice (-2.6 mV, 0.05 nA, 9 nm $\times$ 9 nm).
        }
\end{figure}

\begin{figure*}[t]
        \includegraphics[width=6in]{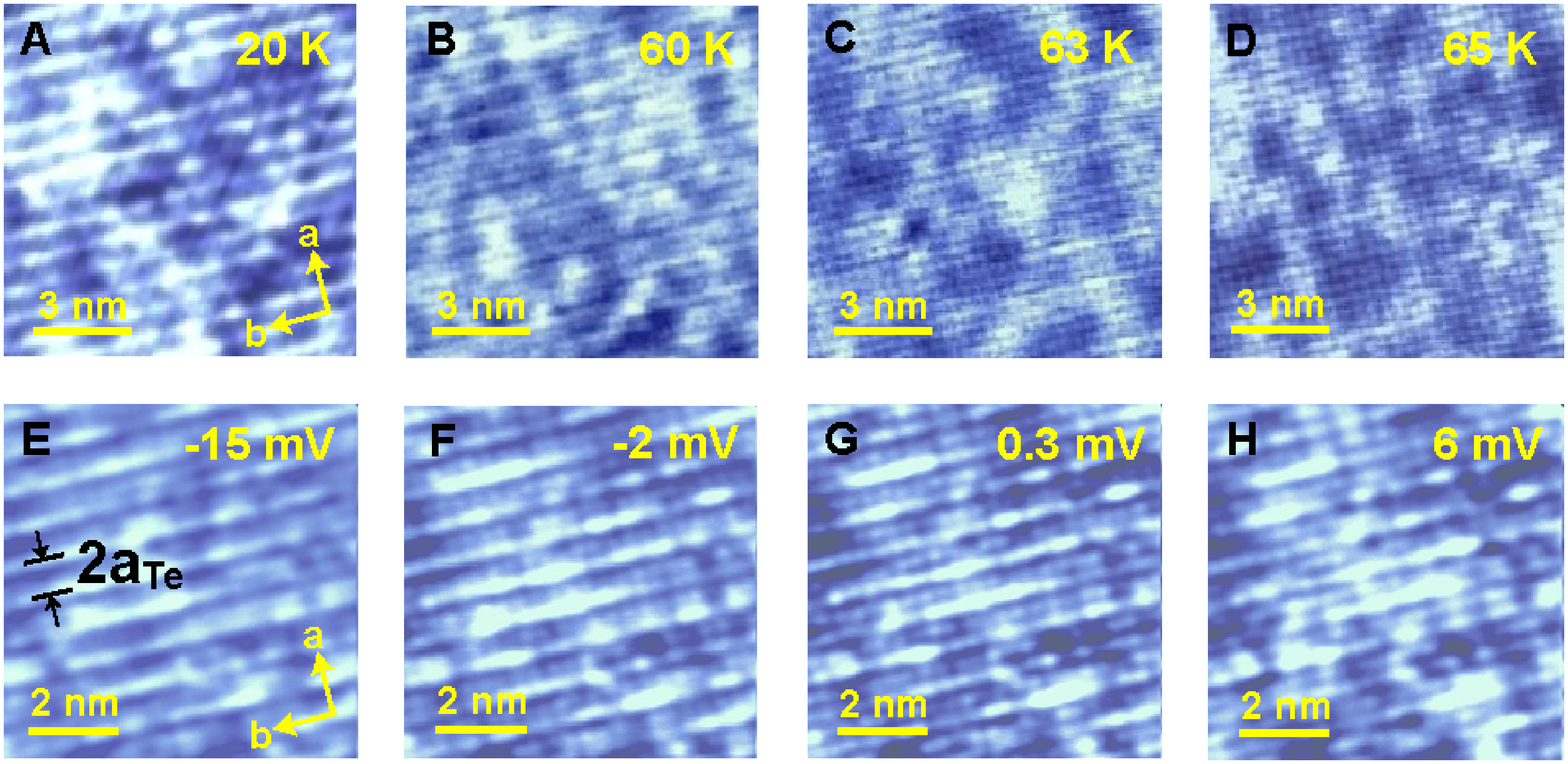}
        \caption{\label{fig2} STM characterization of the charge-stripe order in FeTe.
        (A-D) Temperature-dependence of the surface electronic structure (-5 mV, 0.02 nA, 10 nm $\times$ 10 nm).  Above 65 K, the stripe-like anisotropy completely disappears.
        (E-H) Energy-dependence of the charge order (0.02 nA, 8 nm $\times$ 8 nm). The period is $2a_\mathrm{Te}$ along the AFM direction. The temperature is 4.2 K.}
\end{figure*}

The STM images at different temperatures  (Fig. 2, A to D) shows that the structural transition is correlated with the appearance of a striped pattern.
The pattern is barely visible at 63~K and completely disappears at $T_\mathrm{N} \sim 65$~K.
The stripes are formed by the Te atoms and only visible at very low bias voltage, which rules out the possibility of surface reconstruction.
Measurement of lattice constants by STM identifies the stripe orientation as  the FM direction (b-axis) \cite{SI}.

The energy dependence of the stripes is presented in Fig. 2, E to H, with the STM images of the same location under different bias voltages.
The pattern is only visible at low energy within $\pm30$~meV of the Fermi level. Inside this energy window, the stripes are static and remain unchanged with different  bias voltage.
As clearly demonstrated in the figures,  the bright stripes of Te atoms in the topmost layer alternate with the dark ones in the AFM direction. Therefore the periodicity is $2a_\mathrm{Te}$, the same as  that of the $E$-type spin order. To gain more insight into the origin of the stripes, we mark two adjacent Te atoms in Fig. 1B as  ``A'' and ``B''.  If A is bright, then B must be dark. Here the key point is that these two sites A and B differ in the configuration of their neighboring Fe spins: A is adjacent to one up and three down spins on Fe, while B to one down and three up spins.
Since the charge fluctuations on Fe cations and Te anions are strongly coupled \cite{ma14}, the contrast displayed by the Te atoms actually reflects the existence of a charge-stripe order in the Fe plane. In addition, the concurrence of stripe and AFM order at the structural transition indicates that the charge-stripe order is tied to the long-range $E$-type AFM order.

We note that previous neutron scattering studies \cite{Igor} revealed the existence of strong spin fluctuations in the FeTe system. It is shown that the long-range $E$-type AFM order only makes a small and very low-energy contribution to the entire spin dynamics, while the higher energy part is governed by other competing orders such as the $2\times2$ plaquette spin order \cite{Igor,yin12}. Thus, the correlation between CO and AFM helps explain the appearance of stripes at low energy.

\begin{figure}[t]
        \includegraphics[width=6in]{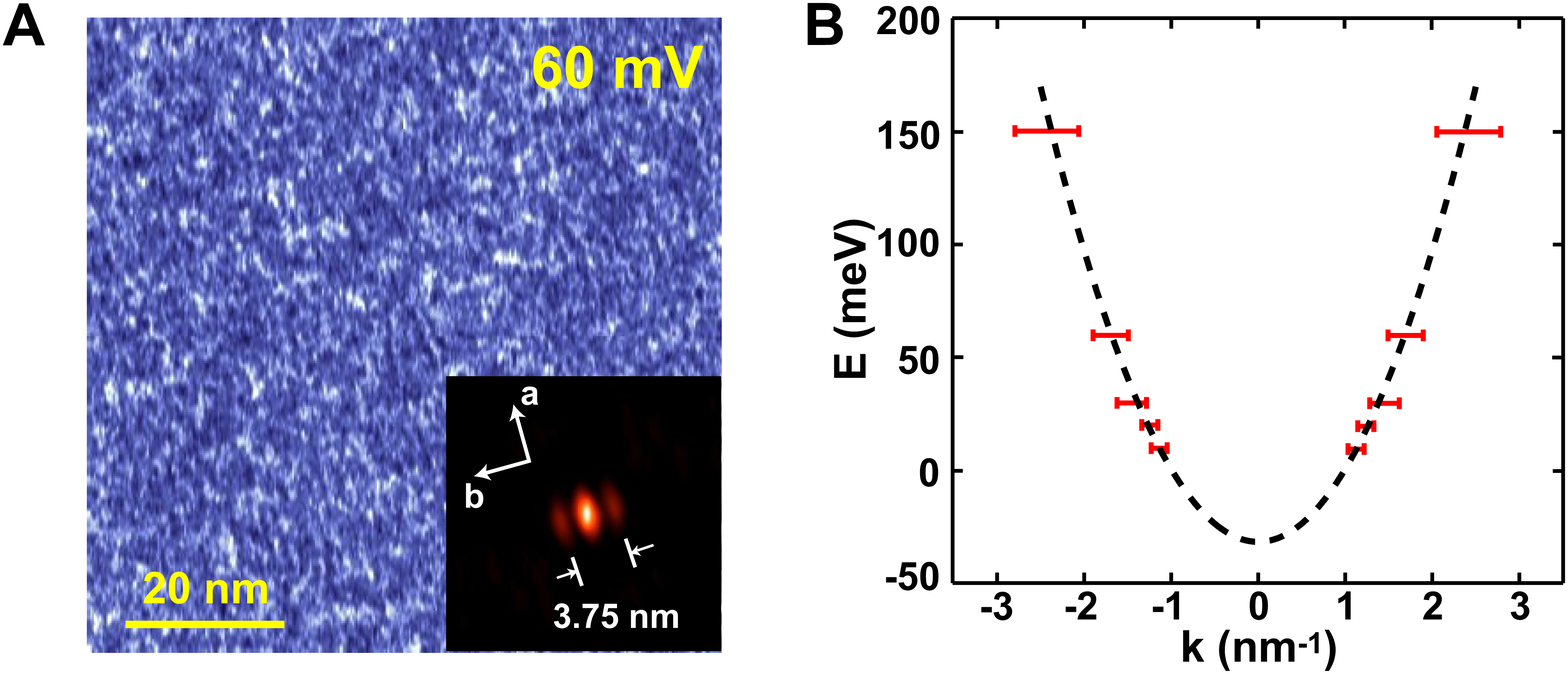}
        \caption{\label{fig:dispersion} Dispersive electronic states along the FM direction.
        (A) $dI/dV$ mapping at 60 mV (0.1 nA, 80 nm $\times$ 80 nm).
         The scattering wave vector is determined by the auto-correlation analysis (see the inset).  (B) Dispersion of the itinerant electrons. }
\end{figure}

Another relevant finding is that the electrons are much more itinerant in the FM direction than in the AFM direction.
An anisotropic dispersion is revealed by scanning tunneling spectroscopy (STS) measurement.
The STS detects the differential tunneling conductance $dI/dV$,
which gives a measure of the local density of states (LDOS) of electrons at energy $eV$.
The $dI/dV$ mapping was performed on the surface of FeTe. At each data point, the feedback
was turned off and the bias modulation was turned on to record $dI/dV$. This procedure resulted in a series of spatial
mapping of LDOS at various bias voltages. A typical $dI/dV$ mapping at 60 mV is shown in Fig. 3A.
The autocorrelation analysis (inset of Fig. 3A) reveals a wave vector exclusively along the b-axis.
A parabolic dispersion (Fig. 3B) is obtained by plotting the wave vector versus energy.
The dispersion is highly anisotropic and only observed in the FM direction. The parent compound of iron-pnictide superconductor Ca(Fe$_{1-x}$Co$_x$)$_2$As$_2$ also exhibits similar property \cite{davis10}. However, the present results for FeTe are nontrivial because of the different electron transport behavior in FeTe from that in the iron pnictides.  In FeTe, the electric conductivity in the FM direction is larger than that in the AFM direction \cite{feng13}, while the opposite is observed for iron pnictides \cite{feng13,fisher10}. The consistency between the anisotropy in band dispersion and electric conductivity in FeTe unambiguously indicates that the itinerant electrons hop much more easily along the FM direction.

The highly anisotropic itinerancy is a reminiscence of the electron behavior in the manganites \cite{hotta03} and suggests that the itinerant electrons in FeTe move in a localized spin background and are favorably described by the double-exchange mechanism in the spin-fermion model \cite{yin10}. In this model, the itinerant electrons and the localized spins $S$ are coupled by the strong on-site Hund's rule coupling $K$, which is a FM exchange interaction. Therefore, an itinerant electron hopping between two sites with the same (opposite) localized spins will experience zero ($KS$) energy barrier, leading to dispersive (nondispersive) electronic states along the FM (AFM) directions and the so-called double-exchange ferromagnetism \cite{anderson55}. The metallic $E$-type AFM spin order in FeTe itself results from a compromised interplay between the double-exchange ferromagnetism and the antiferromagnetism that originates from the superexchange between the localized spins \cite{yin10}.

The observed charge-stripe order in FeTe can be readily reproduced by inclusion of long-range intersite Coulomb interactions $V_{ij}$ (thanks to the poor screening in FeTe) into a model $H_{V=0}$ that can account for the metallic $E$-type AFM spin order with highly anisotropic dispersion.
The Hamiltonian reads  $H=H_{V=0}+H_V$, where
\begin{equation}
H_V=\sum\limits_{ij} { V_{ij} n_i n_j}.
\end{equation}
Here $n_i$ is the electron number operator on the $i$th Fe site. If $m$ refers to the $m$th-neighbor, one may use $V_m=1/\epsilon r_m$ to approximate the screened Coulomb potential, where $\epsilon$ is the dielectric constant of the material and $r_m$ the $m$th-neighbor Fe-Fe bond length. Then the Fe square sublattice leads to $V_1:V_2:V_3=2:\sqrt{2}:1$.

We first analyze the effect of $H_V$ alone without considering the kinetic energy.  Let $\tilde{S}_i^z=n_i-\langle n_i \rangle$, where $\langle n_i \rangle$ is the averaged filling of the itinerant electrons per Fe site. Then, the problem can be mapped to a classical spin model $H_V\equiv\sum_{ij} {V_{ij} \tilde{S}_i^z  \tilde{S}_j^z}$ plus a constant. Substituting $V_m$ by $J_m$, we arrive at the $J_1$-$J_2$-$J_3$ spin model, which is known to yield the $E$-type AFM ``spin'' order when $J_2>J_1/2$ and $J_3>J_2/2$ \cite{xiang09}. This numerical condition is satisfied by $V_1:V_2:V_3=2:\sqrt{2}:1$. Hence, the $V_1$-$V_2$-$V_3$ model alone is ready to yield the observed $E$-type \emph{charge} order with $n_i=2\langle n_i \rangle$ on the ``spin''-up sites and $n_i=0$ on the ``spin''-down sites.

The above charge-stripe order is expected to be weakened by the kinetic energy contributed from $H_{V=0}$. However, the negligible kinetic energy along the AFM direction compared with that along the FM direction warrants the Stoner-type instability of ``spin'' polarization, namely the charge difference $\Delta n$ between the spin-up Fe site and the spin-down Fe site. Therefore, the charge-stripe order still survives even for small $V_{ij}$.  The situation is explicitly shown in fig. S2, where $\Delta n$ is calculated using the aforementioned spin-fermion model for $H_{V=0}$. In the supplemental material~\cite{SI}, we also demonstrate how the effect of local Coulomb interactions is suppressed by the double-exchange mechanism, which is necessary for $V_{ij}$ to be effective in driving the CO instability.

\begin{figure}[t]
        \includegraphics[width=6in]{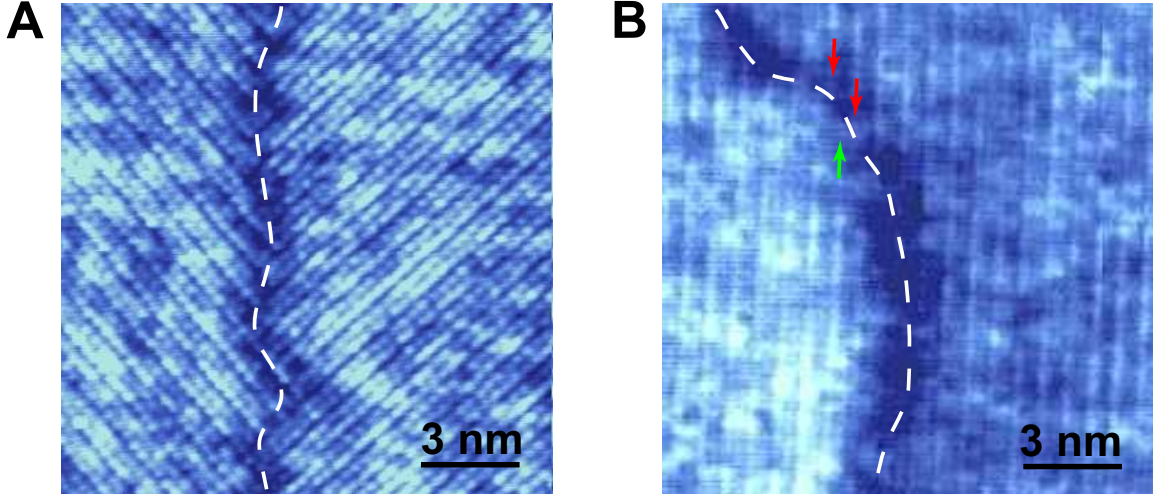}
        \caption{\label{fig:domain}  Stripes across the domain boundaries.
        (A) Twin boundary. The direction of stripes rotate by 90$^\circ$. (B) Anti-phase boundary. Half period is shifted  (indicated by the arrows) across the boundary.
        The dash lines in both images highlight the boundaries. The bias voltages for the two images are 0.2 mV and $-5$ mV, respectively.
        For both images, the scanning size is 15 nm $\times$ 15 nm, and tunneling current is 0.03 nA.}
\end{figure}

Finally, the stripe structure of FeTe indicates two types of domain boundaries: orthorhombic twin boundary and anti-phase boundary.
Both of them have been observed and are shown in Fig. 4. The continuity of the charge stripes ends at the domain boundaries (marked by the dash lines). The stripes either rotate by 90$^\circ$ (Fig. 4A) or shift by $a^{}_\mathrm{Te}$ (Fig. 4B) upon crossing the boundary while the ($1\times1$)-Te lattice in the topmost layer remains uninterrupted. This rules out the effects of possible defects (for example, excessive Fe atoms) as the cause of the observed electronic nematicity.

The $E$-type charge-stripe order revealed in stoichiometric FeTe with integer nominal valence affirms the ubiquity of the phenomenology of CO in SCESs, and suggests a more complex charge dynamics and pairing mechanism than originally expected. It implies that the effect of local Coulomb interaction can be surpassed by the poorly screened longer-range Coulomb interactions in FeSCs, and that bad metallicity, strong correlation from Hund's rule coupling, and strongly coupled charge multipole polarizations of Fe and anion are key to understanding FeSCs.

\clearpage

\pagestyle{plain}

\renewcommand\thefigure{S\arabic{figure}}
\renewcommand\thetable{S\arabic{table}}
\renewcommand\theequation{S\arabic{equation}}
\renewcommand\refname{References}

\setcounter{page}{1}\setcounter{figure}{0}\setcounter{table}{0}\setcounter{equation}{0}

\pagestyle{plain}
\makeatletter
\renewcommand{\@oddfoot}{\hfill\bf\scriptsize\textsf{S\thepage}}
\renewcommand{\@evenfoot}{\hfill\bf\scriptsize\textsf{S\thepage}}
\renewcommand{\@oddhead}{\large\textsf{W.~Li \textit{et~al.}}\hfill\textsf{Supporting Material}}
\renewcommand{\@evenhead}{\large\textsf{W.~Li \textit{et~al.}}\hfill\textsf{Supporting Material}}
\makeatother

\normalsize

\begin{center}
{\vspace*{0.1pt}
\large{Supporting Materials for\smallskip\\\sl
\textbf{\hspace{1pt}Observation of Long-Range Charge Order in Stoichiometric FeTe Films}\smallskip\\
by W. Li, W.-G. Yin, L. L. Wang, K. He, X. C. Ma, Q.-K. Xue, and X. Chen}}
\end{center}

\section{Materials and Methods}

The FeTe (001) film was prepared on the graphitized 6H-SiC (0001) substrate. High-purity Fe (99.995\%) and Te (99.9999\%) were evaporated from two standard Knudsen cells. The growth was performed in the Te-rich conditions with a nominal Te/Fe flux ratio of $\sim$15 to avoid excess Fe in the film, while the substrate temperature was held at 310$^\circ$C. The growth follows the typical layer-by-layer mode. The as-grown films were directly transferred to STM.
A polycrystalline PtIr STM tip was used in the experiments. The STM topographic images were processed using WSxM (www.nanotec.es). We studied the samples with the thickness of $15$, $20$, and $30$ unit cells and found similar results. We present here the measurements on the $30$-unit-cell sample.

\section{Supporting Online Text and Figures}
\subsection*{Identification of the FM and AFM directions from topographic images}

The image with atomic resolution in Fig. S1 exhibits a quasi-square lattice of Te atoms with lattice constant of $a^{}_\mathrm{Te}\simeq$ 3.8~\AA. Previous studies \cite{Sbao09,Sdai09} on bulk FeTe show a tetragonal to monoclinic (approximately orthorhombic) structural phase transition at $T_\mathrm{s} \sim 65$~K with a simultaneous development of the $E$-type antiferromagnetic (AFM) order (i.e., $T_\mathrm{N}=T_\mathrm{s}$); the long axis of the Te lattice is along the AFM direction (the a-axis in Fig.~1B). To determine the lattice orientation of the film, we performed the Fourier transform of Fig.~S1 (see the inset of Fig.~S1). The lattice constant along the a-axis is 2\% larger than that along the b-axis. We therefore identify the a-axis as the AFM direction and b-axis as the ferromagnetic (FM) direction in the film.

\begin{figure}[h]
\begin{center}
        \includegraphics[width=3in]{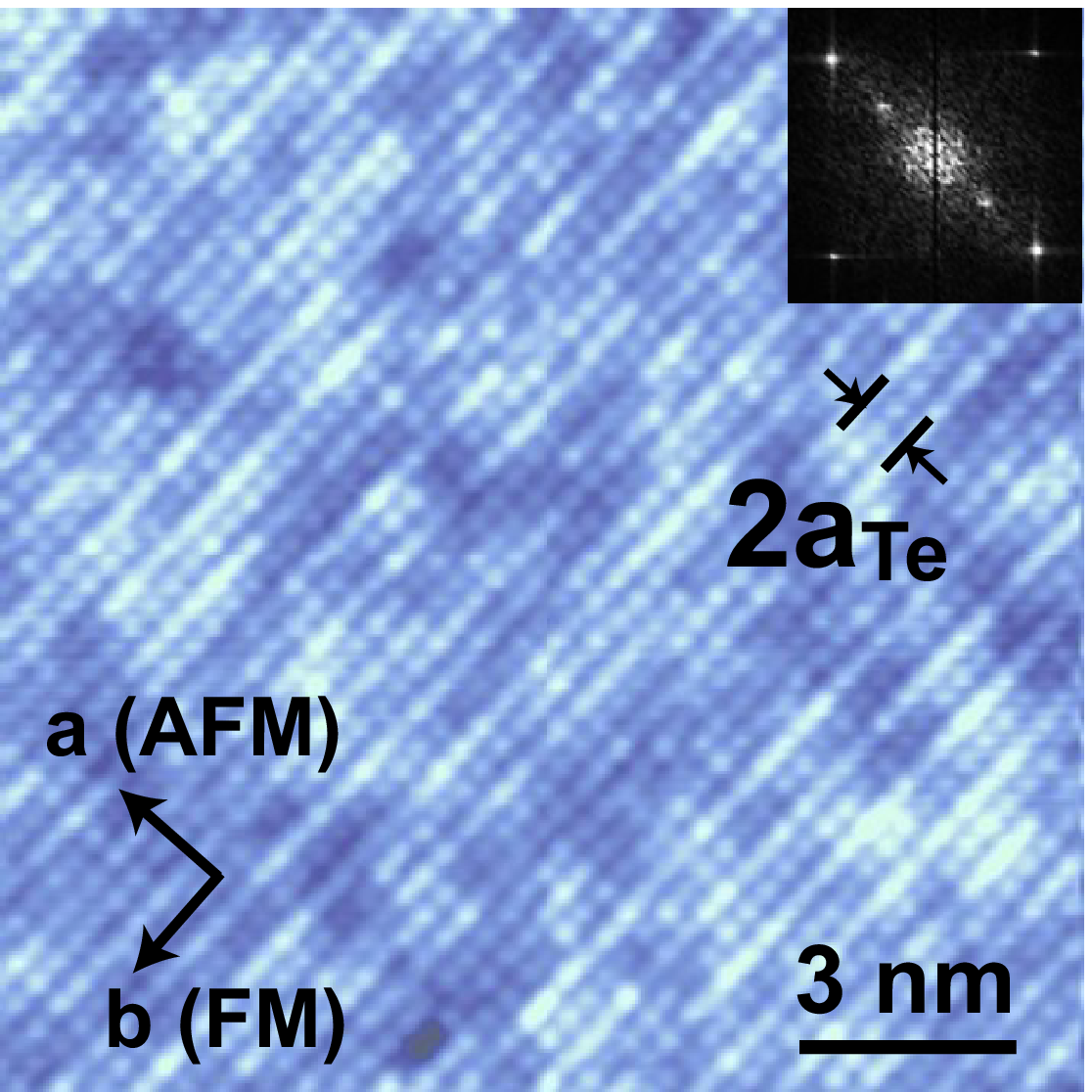}
        \caption{\label{fig1_ab}  Atomic resolution STM topography of FeTe film (1.5 mV, 0.08 nA, 15 nm $\times$ 15 nm).
        Inset is the Fourier transformation, from which the FM and AFM directions are identified.}
        \end{center}
\end{figure}

\subsection*{Calculation of charge order in the spin-fermion model}
The effective Hamiltonian becomes $H=H_\mathrm{S-F}+H_V$. The spin-fermion model $H_\mathrm{S-F}$, where the Fe $d_{xz}$ and $d_{yz}$ orbitals were treated to host itinerant electrons and the rest Fe $3d$ orbitals were treated as an effective localized spin, was proposed as a minimal model to unify the various magnetic orders observed in FeSCs, such as metallic \textit{E}-type in FeTe, metallic \textit{C}-type in LaOFeAs and BaFe$_2$As$_2$ \cite{Syin10,Syin12_SST} and insulating $2\times 2$ block-type in K$_2$Fe$_4$Se$_5$ and BaFe$_2$Se$_3$ \cite{Syin12}, $A$-type in TaFeTe$_3$ \cite{Ske12,Sxu14}, and FM in CuFeSb \cite{Sqian12}, etc. It reads \cite{Syin10,Syin12_SST,Slv10,Sweng09,Svishwanath09,Sliang12}
\begin{eqnarray}
\label{eq1} H_\mathrm{S-F}=&-&\sum\limits_{ij\gamma \gamma^\prime \mu }
{(t_{ij}^{\gamma \gamma^\prime} C_{i\gamma \mu }^\dag C_{j
\gamma^\prime \mu}^{} +h.c.)} 
- \frac{K}{2}\sum\limits_{i\gamma \mu \mu' } {C_{i\gamma \mu
}^\dag \vec {\sigma }_{\mu \mu' } C_{i\gamma \mu' }^{} }  \cdot \vec
{S}_i + \sum\limits_{ij} { J_{ij} \vec {S}_i \cdot \vec {S}_j} \nonumber \\
& + & U\sum\limits_{i\mu} {n_{i\gamma \uparrow}^{} n_{i\gamma \downarrow}^{} }
  +  U'\sum\limits_{i\mu \mu'} {n_{i,xz,\mu}^{} n_{i,yz,\mu' }^{} }
  - J_H\sum\limits_{i\mu} {n_{i,xz,\mu}^{} n_{i,yz,\mu}^{} },
\end{eqnarray}
where $C_{i\gamma \mu }^{} $ denotes the annihilation operator of an
itinerant electron with spin $\mu=\uparrow$ or $\downarrow $ in the
$\gamma=d_{xz}$ or $d_{yz} $ orbital on site $i$. $t_{ij}^{\gamma
\gamma^\prime} $'s are the electron hopping parameters. $\vec
{\sigma }_{\mu \mu' } $ is the Pauli matrix and $\vec {S}_i$ is the
localized spin whose magnitude is $S$. $K$ is the effective Hund's rule coupling between the itinerant electrons and the localized spins. $J_{ij}$ is the AF
superexchange couplings between the localized spins; in particular,
$J$ and $J'$ are respectively the nearest-neighbor (NN) and
next-nearest-neighbor (NNN) ones. The $U$, $U'$, and $J_H$ terms on the second line of Eq.~\ref{eq1} describes the on-site intraorbital Columbic, interorbital orbital Columbic, and Hund's rule interactions between the itinerant electrons, respectively. The filling of the itinerant electrons is three per Fe site, corresponding to the high-spin configuration of Fe $3d^6$ \cite{Syin09}.

To minimize the number of free parameters, we take $U$ and $V_3$ as free parameters, set $V_1:V_2:V_3=2:\sqrt{2}:1$ and $U'=U-2J_H$, and keep the other parameters the same as previously used for FeTe \cite{Syin10}, notably $KS=J_H=0.8$ eV. We also keep the same treatment of the localized spins as Ising spins, which has been shown to suffice for the problem of interest.

In Fig. S2, we present the calculated $\Delta n$ as a function of $V_3$ for three typical cases: (i) $U=U'=J_H=0$ (black solid line), (ii) $U=2.4$ eV, $U'=0.8$ eV, and $J_H=0.8$ eV (red dotted line), and (iii) $U=4$ eV, $U'=2.4$ eV, and $J_H=0.8$ eV (blue dashed line). It shows that as $V_3$ increases, the $E$-type CO saturates to $\Delta n = 2\langle n_i \rangle$ as discussed in the main text. As $U$ increases, the CO is suppressed considerably as expected. The results for the first two cases are comparable to each other. This is understandable as the double-exchange effect: The strong $KS$ term tends to align the spins of the itinerant electrons with the localized spins, thus suppressing the effect of $U$ and the effective interaction between the itinerant electrons becomes $U'-J_H$, which is zero in case (ii). This justifies the use of the noninteracting case (i) to approximate the interacting case (ii) \cite{Syin10,Syin12_SST,Syin12}. We show that the CO starts to appear at small $V_3\simeq 0.1$~eV.

\begin{figure}[h]
\begin{center}
        \includegraphics[width=4in]{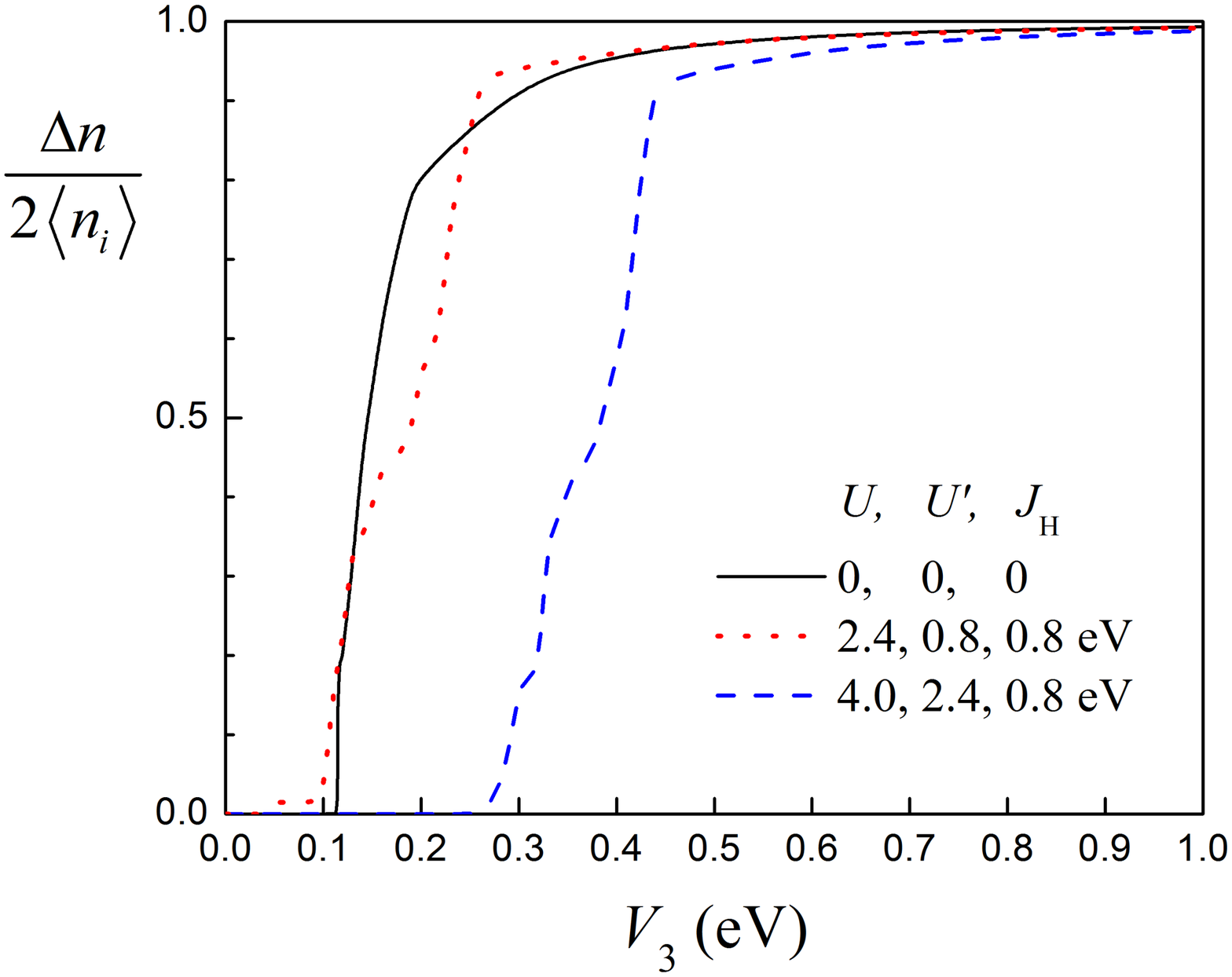}
        \caption{\label{fig:theory} The charge difference $\Delta n$ between the spin-up Fe site and the spin-down Fe site as a function of $V_3$ calculated from using the spin-fermion model \cite{Syin10,Syin12} with an addition of $V_1:V_2:V_3=2:\sqrt{2}:1$ for three typical cases.
        }
                \end{center}
\end{figure}


\begin{thebibliography}{99}
\bibitem{note:bad-metal}
Bad metals are referred to materials whose electric resistivity shows metallic-like temperature dependence but large (usually of the order of m$\Omega$ cm) absolute value and whose optical Drude weight is finite but small.
\bibitem{SI} Materials and methods are available as online supporting material.
\end{thebibliography}
\end{document}